\documentclass[12pt]{article}
\usepackage {amssymb}
\usepackage {amsmath}
\usepackage{graphicx}

\begin{document}

{\footnotesize{\noindent in {\it Developments in Quantum Physics},
eds. F. Columbus and V. Krasnoholovets (Nova Science Publishers
Inc., New York, 2004), pp. 85-109}}

\begin{center}
\section* {\textbf{On the Origin of Conceptual Difficulties of
\break Quantum Mechanics}}
\end{center}

\medskip\bigskip

\begin{center}
{\large {\bf Volodymyr Krasnoholovets}}
\end{center}

\medskip

\begin{center}
{Institute of Physics, National Academy of Sciences \\ Prospect
Nauky 46, UA-03028 Ky\"{\i}v, Ukraine}
\end{center}

\medskip

\hspace{10 cm} {23 January 2003}

\begin{center}
{\bf Abstract}
\end{center}

{\small It is the matter of fact that quantum mechanics operates with
notions that are not determined in the frame of the mechanics'
formalism. Among them we can call the notion of "wave-particle"
(that, however, does not appear in both classical and high energy
physics), the probabilistic interpretation of the Schr\"odinger
wave $\psi$-function and hence the probability amplitude and its
phase, long-range action, Heisenberg's uncertainty principle, the
passage to the so-called operators of physical values, etc.
Orthodox quantum mechanics was constructed as a physical theory
developed in the phase space of the mentioned notions. That is why
the formalism of quantum mechanics is aimed only at detailed
calculations of the stationary states of the energy of the quantum
system studied and is not able to describe a real path running by
the system in the real space; instead, the formalism gives an
averaged probabilistic prediction. Thus, if we are able to develop
quantum mechanics in the real space, an option to clarify all the
difficulties associated with the above notions would appear. Such
a theory of quantum mechanics developed in the real space in fact
has recently been constructed by the author. The theory started
from deeper first principles, namely, from the consideration of
the notion of a 4D space-time. So, the notion of fundamental
particle, the principles of the motion of a particle and other
characteristics have been made clear. The theory, rather a
submicroscopic one, is characterized by short-range action that
automatically means the introduction of a new kind of carriers,
i.e. carriers of the quantum mechanical force. The existence of
the carriers called "inertons" (because they carry inert
properties of matter) has indeed been verified in a number of
experiments.

\bigskip
\bigskip

{\bf  Key words:} \ \  quantum mechanics, space, matter waves,
inertons
\\

{\bf PACS:} \ \  03.65.Bz Foundations, theory of measurement,
miscellaneous theories; \ 03.65.w Quantum mechanics; \ 03.75.-b
Matter waves; \ 14.80.-j Other particles (including hypothetical)

}

\newpage

\section{Dominant views on the foundations of quantum mechanics}

\hspace*{\parindent} The founders of quantum mechanics raised some
problems regarding its conceptual difficulties such as the
description of particles by using a probabilistic wave
$\psi$-function, the $\psi$-function collapsing at the measuring,
Bohr's notion of "complementary" [1], Schr\"odinger's cat (i.e.,
the problem of transforming a set of microscopic states into a
particular macroscopic state) and entanglement [2],
Einstein-Podolsky-Rosen (EPR) paradox [3], etc. Since that the
subject still continue to disturbance inquiring minds of
researchers and nowadays they put the study on nonlocality of
quantum mechanics, quantum paradoxes, entanglement and
decoherence, and quantum teleportation in the forefront of the
fundamental science.

Among recent research on misconceptions regarding quantum
mechanics one can mention the paper by Bush [4] in which he
touches a number of conceptual and mathematical problems in the
fundamentals and the study by Styer [5] in which he lists a series
of commonly held misconceptions such as "energy eigenstates are
the only allowed states", "the wave is dimensionless", etc. It
seems that the literature on the so-called Bell inequality [6] and
its validity or violation is the most extensive. Evidently that
such an interest to the Bell inequality is caused by the fact that
Bell's inequality is based on other postulates of the physical
world than orthodox quantum mechanics prescribes.

    Let us touch studies regarding the Bell theory in some detail.
Muynck [7] noted that the Bell inequality was first derived for
hidden-variables theories, rather stochastic theories (see, e.g.
Refs. [8,7] on the hidden-variables theories) and only after that
the inequality was obtained for conventional quantum mechanics.
Many years ago Wigner [9] considering the problem of hidden
variables also noted that we could not obtain directly their
magnitudes. He noticed that the measuring the distribution of
magnitudes of hidden variables still remained undetermined, or
fuzzy. Therefore, hidden variables as such severe suffer from a
statistical nature.

Let us write the Bell inequality following Muynck's designations
[7]. In the Bell experiment either standard observable $A_1$ or
$B_1$ of particle 1 is measured jointly with standard observable
$A_2$ or $B_2$ of particle 2. It is assumed that observables $A_1$
and $B_1$ are incompatible, i.e. commutator $[A_i,B_j]\neq 0$,
though $[A_i,B_i] = [A_j,B_j] = 0$ where $i=1, {\kern 1pt} 2$,
$j=1, {\kern 1pt} 2$, and $i \neq j$. Let all observables have
only values $+1$ and $-1$. Then in terms of the expectation values
$\langle A_i A_j \rangle$ of the correlation observables $A_i A_j$
the Bell inequality becomes
\begin{equation}
\big|  \langle A_1 A_2 \rangle -  \langle A_1 B_2 \rangle  \big|
\leq 2+ \langle B_1 A_2 \rangle +  \langle B_1 B_2 \rangle.
\label{1}
\end{equation}

Mensky [10] emphasized such major peculiarities of the objective
local theory by  Bell as: (i) each particle is characterized by a
number of variables (in expression (1) two variables are analysed)
which are possibly correlated for the two particles; (ii) the
results of measurement of one particle do not depend on whether
the other particle is measured or not, and if it is, they do not
depend on the result of such a measurement. Thus the Bell
experiment does not suggest any quantum mechanical nonlocality.

Experimental data by Aspect and co-workers [11,12], Zeilinger's
team [13] and others showed the violation of Bell's inequality
(1). Many researchers were astonished at such a result. In
particular, Evdokimov et al. [14] noted that the violation of
Bell's inequality in quantum world was a true paradox.

Muynck [7] accounts for the Aspect's experiments by a local
disturbance of the measurement results due to incompatibility of
observables of the same particle, rather than of a nonlocal
disturbance of one particle due to a measurement of a observable
of the other particle. Muynck analyzes also the generalized
inequalities and their violation by the experiments in connection
with the experiments. He reasonable notes that we do not have any
reason to expect that the standard Aspect experiments should
satisfy the Bell inequalities, because of Heisenberg disturbance
of a complicated probability. In other words, the violation of
Bell's inequality is a consequence of the observables, which
definitely should be caused by Heisenberg's uncertainty principle
that occurred in the arms of the interferometer. So, there is no
need to suppose any extension nonlocality, i.e. the influence of
the measurement in one arm of the interferometer on the
measurement in the other one.

Nakhmanson [15,16] directly shows weak points in the experiments
by Aspect's and Zeilinger's teams. He notes that the actual
experiments were far from the thought experiment by EPR and Bell,
which the experimenters washed to test. Namely, in the Aspect's
case the switching of conditions of registration of photons was
not adiabatic and therefore it prevented any realistic connection
between photons in the EPR pair. Nakhmanson wrote: "This gave rise
to the legend of nonlocality of quantum mechanics, of the
'instantaneously' correlated behavior of the EPR pair, even though
its constituent particles may be hundred of light-years apart".
Then he continued that in the experiments by Zeilinger's team [13]
the conditions of registration were governed by a random number
generator and the 'randomness' was borrowed from the object of the
study itself, i.e. the quantum world.

Eberly [17] has recently argued that the inequalities of the Bell
type by themselves have nothing to do with quantum theory. He
points out that the inequalities do not take into account that
there is no physical sense to an intermediate polarization in
quantum theory. However, that is that state that Bell's
inequalities have included. Then Eberly stresses: "The violation
of Bell's inequality is a simple mathematical result obtained by
an uncomplicated counting of members of objects in clearly defined
categories."

Nevertheless, the study of Bell's inequalities still continue.
Mermin [18] has recently shown that the some kinds of correlation
(or even direct classical communication) between detectors in fact
validate Bell's theorem. Golshani and Fahmi [19] asserts that the
violation of Bell's inequality is not necessarily the violation of
Bell's locality condition and that if there is any nonlocality in
nature, it is not in the form of even more complicated
inequalities studied by the authors. The most detailed study
demonstrates Zeilinger's team (see, e.g. Refs. [20-24]). They
introduce, in particular, the notion of 'quantum entanglement'
determining it as a feature of a composite system to have more
information contained in correlations than any classical mixture
of its individual constituents could ever have [22]. Working on
the developing of the quantum theory of information they have
proposed [23] the notion of the irreducible randomness of
individual events, which together with quantum complementary and
quantum entanglement have allowed them to support the nonlocality:
The nonlocality for photons that never interacted has been
confirmed by observing a violation of Bell's inequality by 4.5
standard deviations [24]. The result obtained is considered as an
actual proof of the quantum nature of teleportation. Thus
Zeilinger's team completely supports Bohr's idea that they quoted
in Ref. [23]: "There is no quantum world. There is only an
abstract quantum physical description. It is wrong to think that
the task of physics is to find out how Nature is. Physics concerns
what we can say about Nature."

In interesting theoretical work [25] interference and entanglement
have recently been studied in the frame of a so-called intrinsic
approach. As a rule classical states are prescribed to the phase
space, while quantum states, which are considered as linear
functions on the dynamical variables, assign to the vector space
formalism (the Hilbert space). Density matrices belong to
bilinears in the vectors. The authors [25] propose a generalized
method of restoring an entangled pure state (a purification of the
linear vector) from mixed states of the subsystems involved in
entanglement. The pure density matrix is then treated as a measure
of entanglement for the quantum system in question. In orthodox
quantum mechanics linear operators, which act on the Hilbert space
of states, are associated with observables. For instance, the
Wigner distribution $f(\bf x)$ is related to observables as well.
Nevertheless, although $f(\bf x)$ is defined in the Hilbert space
[25] its arguments belong to the phase space (note in the general
case the Wigner function is determined as $f({\bf x}, \; {\bf p},
\; t )$). So the phase space and the Hilbert one appear as
interconnected, though the Hilbert space describes strictly
stationary states of the quantum system in question.

Stapp [26,27] analyzing both theoretical and experimental results,
which touch questions like these: Is quantum theory local or
nonlocal? and Is nonlocality is real?, has adduced many arguments
for this or that point of views and specifically noted that
quantum theory is still formulated as an indeterministic theory.

Among other urgent topics that have been developing we would like
to mention research by Cramer [28], Griffiths [29], and Holland
[30]. Cramer [28] considered an explicit and fully causal nonlocal
mechanics for describing quantum events, which allowed him to
claim about new insight into the reality behind the quantum
mechanical formalism. The basic element of his theory is
transaction describing a quantum event as an exchange of advanced
and retarded waves. However, the author did not touch a nature of
the waves, i.e. whether they are electromagnetic, inert,
gravitational, or some other lineage. Moreover, he noted that the
verification of the theory met grave difficulties.

Griffiths [29] introduces probabilities and stochastic processes
as part of the foundations of quantum mechanics. He uses the
mathematics of Hilbert space and constructs quantum principles
known as consistent histories, or decoherent histories, that
appears in his approach as basic principles of quantum theory.

Holland [30] adventures causal deterministic ideas by de Broglie
and Bohm, though a surrounding quantum wave that guides a particle
is treated in the framework of statistical mechanical type
probabilities. For further reading on the Bohmian mechanics we
would like to refer the reader to Goldstein [31]. Some other
interpretations of quantum mechanics are stated in Refs. [32,33];
some aspects of the measurement problem and approaches to its
solution are posed in Ref. [34].

Although quantum mechanics demonstrates a unique predictive
success, it is still hampered by severe conceptual difficulties
that scientists have tried to clarify. Looking for the answer to
the question: What is quantum mechanics about?, researchers extend
frames of the probabilistic basis of quantum theory. Nevertheless,
quantum mechanics as such seems as a quite closed theory. Mermin
[35] holds the same opinion: "First of all, by 'quantum mechanics'
I mean quantum mechanics as it is -- not some other theory in
which the time evolution is modified by nonlinear or stochastic
terms, nor even the old theory augmented with some new physical
entities (like Bohmian particles) which supplement the
conventional formalism without altering of its observable
predictions. I have in mind ordinary everyday quantum mechanics."
Muynck [36] shares this view as well: "Whether quantum mechanics
just describes the 'phenomena', or whether it describes 'reality
behind the phenomena'? In particular, within the quantum mechanics
formalism I find no reason to conclude that experimental violation
of the Bell inequality would imply any nonlocality (even though
there is some nonlocality  involved in the quantum mechanical
formalism!). In my view the conceptual difficulties of quantum
mechanics are mainly man-made, as a consequence of too high an
expectation of the extent to which this theory describes 'reality
behind the phenomena'."

Notwithstanding this, the inquisitive mind wishes to light upon a
secret 'behind the phenomena.' In what way can it be done? It is
obvious that we should extend quantum mechanics to the sub atomic
area at which the initial notions quantum mechanics operates with
will become clear. Among such notions we first of all should
mention the following: de Broglie wavelength, Compton wavelength,
wave-particle, $\psi$-function, spinor, operators of physical
values, commutativity and noncommutativity of operators,
Heisenberg's uncertainties, spin, Pauli exclusion principle,
Lorentz noninvariance of the Scr\"odinger equation,  discrepancy
between Schr\"odinger's (nonrelativistic) and Dirac's
(relativistic) formalisms and the absence of an intermediate
formalism, Zitterbewegung, and so on.

If we are capable to make clear all of the basic notions of
quantum theory, we in fact will clue the constitution of Nature at
a deeper level. On the other hand, disclosing the notions above we
shall automatically arrive at the complete determinism, that is,
submicroscopic deterministic quantum theory... Such a theory being
constructed will unravel new peculiarities and links in quantum
systems (including those that fall within electroweak and strong
interactions).

\section{Clarification of fundamental notions}

\subsection{Space}
\hspace*{\parindent} De Broglie [37] wrote about a subquantum
medium that should be presented in works studying the problem of
the causal interpretation of quantum mechanics. But what kind of a
medium? Many of his ideas disclosed Lochak (see, e.g. Ref. [38]),
the nearest collaborator of de Broglie and the President of
Fondation Louis de Broglie (Paris, France). Nowadays we face a
huge number of works dedicated to such kind of a medium that many
researchers call an aether. Researchers who construct models of an
aether note that conceptual difficulties of orthodox quantum
theories rise just from denying an aether, because doing so we
exclude the possibility of matter to interact with the aether.
Some of these researchers see all particles as being conceived
from a unique corpuscle (see, e.g. Ref. [39,40]) and, moreover,
some direct experimental evidences of the interaction of matter
with a subquantum medium is indeed demonstrated by Baurov [40].

What was the opinion of Einstein about an aether? In his
well-known work "Aether and the theory of relativity" [41] he
stated that since space was endowed with physical qualities, an
aether existed. Then he mentioned that according to the general
theory of relativity space without an aether is unthinkable: in
space without an aether light would not propagate; there would not
any space-time intervals in the physical sense, etc. However, in
summary Einstein stressed that this aether might not be thought of
as endowed with quality characteristic of ponderable medium, as
consisting of parts that might be tracked through time.

Notwithstanding Einstein's determination that rejected the idea of
likeness between an aether and a typical medium, quantum
mechanics, which was constructed later, introduced a new
understanding of space. In work "Is there an aether?" Dirac [42]
considering relations between general relativity and quantum
mechanics noted that although the relativity posed the objections
to an aether, quantum mechanics practically removed them. Thus, a
subquantum medium in fact might be intervened in the behavior of
moving particles introducing a peculiar quantum mechanical
strangeness.

However, researchers still did not set themselves the problem of
constructing a detailed theory of quantum mechanics derived from
the structure of space. Hitherto we have met only incomplete and
uncoordinated models and concepts of space and quantum mechanics,
which consider only separate aspects of the structure of space and
the interaction of quantum systems with the space. In no
circumstances such studies cannot be called rigorous theories of
quantum behavior of matter. It should be particularly emphasized
that since the beginning of the 1950s de Broglie's major idea [36]
was the construction of a double solution theory. The theory
should satisfy the Scr\"odinger formalism and simultaneously
introduce the deepest deterministic description of the behavior of
a quantum system.

It seems that so far the author in fact has been along who tried
to follow de Broglie's ideas as close as possible. The main
starting point of the author's concept [43,44] was the inner
construction of a subquantum medium. Why do we need a substrate?
Because a specialist who works in the field of condensed matter
physics cannot see particles that move and interact in an absolute
emptiness, i.e. a vacuum. Note that in the case of condensed
matter, particles move in a lattice of atoms/molecules. The
lattice can be ordered or disordered, dense or not compact, but it
always exists! Besides, experiments in high energy physics
unambiguously demonstrate that particles can be created in any
place of space and, moreover, high energy physics predicts a
critical size, $10^{-30}$ m, at which all kinds of physical
interactions should coincide. Furthermore, high energy physicists
operate with the notion of an abstract superparticle whose
different states are electron, positron, muon, quark ($u$-quark,
$d$-quark, and the others), etc. Thus a fuller picture of the
substrate was beginning to emerge: The primary substrate might be
tightly composed of real superparticles, primary blocks, whose
size is around $10^{-30}$ m.

Those first steps then passed into a rigorous mathematical theory
of the real space developed by Bounias and me in a series of works
[45-47]. Why do we prefer the notion of the space over that of an
aether? The author tried to answer this interesting question in
article [48]. The matter is that this is the pure historical
issue: ancient Indian sages possessed an inexplicable deep
knowledge regarding the constitution of the physical world and
they used term 'space' ('loka' in Sanskrit) [49]; ancient Greeks
partly borrowed their knowledge and transformed 'loka' into
'aether'. It is interesting that Newton also adhered the notion
'space' that was in his opinion constructed of compactly stacked
small rigid balls.

In our works [45-47] studying the constitution of the real space
(i.e., a 4-D space-time) we have used topology, set theory and
fractal geometry. An abstract lattice of empty set cells $\O$ has
been shown to be able to account for a primary substrate in a
physical space. This lattice is a fractal lattice if it allowed
for the magma $\O^{\O}=\{ \O,{\kern 1pt} \complement \}$
constructed with the empty hyperset and the axiom of availability.
Spacetime is represented by ordered sequences of topologically
closed Poincar\'{e} sections of this primary space. We have
demonstrated that the antifounded properties of the empty set
provide existence to a lattice involving a tessellation of the
corresponding abstract space with empty balls. This structure has
thus been called a {\it tessellattice}. M. Bounias introduced the
'Moment of Junction' that allowed us to investigate the
composition of indicative functions of the position of points
within the topological structures and to account for elements of
the differential geometry of space-time.

The tessellattice represents a degenerate space-time, i.e. in this
case all cells (in other words, balls) are degenerated. A
particled ball provides a formalism describing the elementary
particles proposed in Refs. [43,44]. In this respect, mass is
represented by a fractal reduction of volume of a ball, while just
a reduction of volume as in degenerate cells, which was initially
postulated in Refs. [43,44,50-52], is not sufficient to provide
mass (because a dimensional increase is a necessary condition).
Accordingly, if $\rm V_o$ is the volume of an absolutely free
cell, then the reduction of volume resulting from a fractal
concavity is the following: $\rm V^{part} = V_o - V_f$. The mass
$m$ of a particled ball is a function of the fractal-related
decrease of the volume of the ball:
\begin{equation}
m \varpropto (1/{\kern 1pt}{\rm V}^{\rm part}) \cdot ({\rm
e}_{{\kern 0.5pt}v}-1)_{{\kern 1pt}{\rm e}_{{\kern 0.5pt}v}>1}
\label{2}
\end{equation}
where (e) is the Bouligand exponent, and $\rm (e-1)$ the gain in
dimensionality given by the fractal iteration.

The moment of junction allows the formalization of the topological
characteristics of what is called {\bf \textit{motion}} in a
physical universe. It is the motion that was called by de Broglie
as the major characteristic determining physics. While an identity
mapping denotes an absence of motion, that is a null interval of
time, a nonempty moment of junction stands for the minimal of any
time interval. Sidharth [53] argued that a minimum spacetime
interval should exist and that "one cannot go to arbitrarily small
spacetime intervals or points". In our sense, there is no such
"point" at all: only instants that at bottom of fact do not
reflect timely features.

In such a manner, a (physical) vacuum that is hazy something or
nothing in all modern quantum theories should be replaced by the
real space, or physical space stated above.

High energy physics working on sub microscopic scales proposes
some Higgs condensate that would be initial at the creation of the
physical world. Nonetheless, the Higgs condensate of models of
grand unification of interactions is not constructed in a real 4D
space-time and moreover, it does not give any idea in what way it
can manifest itself in quantum mechanics. Inasmuch as quantum
mechanics is the most reliable basis for all other quantum
theories, any new quantum concept has to produce orthodox quantum
mechanics as a limiting case of the theory constructed (i.e. de
Broglie's thesis concerning the double theory solution should be
satisfied). However, either quantum chromodynamics, or some other
contemporary theory such as string theory and others are not able
to mutate in the orthodox quantum mechanical formalism. Quantum
field theories and their derivatives suffer from undetermined
field variables ($\varphi$, $\varphi^4$, etc.) constructed in
abstract spaces, and so on. Group methods also isolate themselves
from both the constitution of the space and the direct
measurement.

General relativity does not deal with any substrate, its major
subject is geometry. However, we should not forget that the
relativity separates the mass from the geometry, i.e. matter from
space. If we assume that matter appears from the space, which in
turn is a substrate, we immediately arrive at the conclusion that
the matter should interact with such a space: the space itself
becomes material.

Thus the submicroscopic theory of the real space constructed in
Refs. [45-47] discloses very new horizons in physics.  In the area
of macroscopic physics this is the possibility of deriving the
theory of gravity starting from quantum mechanics. Recently a
definite success in this direction in fact has been achieved
[50,54-56]. In the sub atomic area the reconsideration of the
strong and electroweak interaction starting from deeper first
principles is still waiting for  pioneers.

\subsection{Long-range action}

\hspace*{\parindent} Ehrenfest [57] pointed out that by
conventional quantum mechanics, particles can interact
simultaneously even if they are spaced at any quantity of
kilometers; he even exclaimed: "What a strange theory we have!"
Long-range action of quantum mechanics was also emphasized by
Pauli [58]; in particular, he noted that quantum mechanics bears
up against a hypothetical basis that the speed of the interaction
in the quantum mechanical range $c=\infty$ and that the
gravitational interaction is negligible, the constant of
gravitational interaction $G=0$.

  For instance, let us turn to the problem of hydrogen atom, a
typical example of long-range action in quantum mechanics, which
also was stressed by Arunasalam [59]. The radial part of the
Schr\"odinger equation written for a particle in a spherically
symmetric electrostatic potential $V(r)$ has the form (see, e.g.
Schiff [60])
\begin{equation}
-\frac{\hbar^2}{2m} \frac {d^{\kern 1pt 2} \chi}{d r^2} + \Big[
V(r) + \frac {l(l+1) \hbar^2}{2mr^2} \Big] \chi = E \chi \label{3}
\end{equation}
where $\chi(r)$ is the radial wave function. The second term in
the square brackets is stipulated by the potential energy
associated with the moment of momentum of the particle. The
potential energy
\begin{equation}
V(r) + \frac {l(l+1) \hbar^2}{2mr^2} \label{4}
\end{equation}
ensures the stability of the particle orbit. In the case of the
hydrogen atom the potential $V(r) = e^2/(4\pi \epsilon_0{\kern
1pt} r)$ and the equation of related motion of an electron and
proton has the form similar to Eq. (3).

However, it should be noted that the Schr\"odinger {\it quantum}
equation (3) includes the potentials $V(r)$ written in pure {\it
classical} {\kern 0.5pt} terms, much as in the problem of Newton
gravity where $V(r)=G m/r$! In Eq. (2) $V(r)$ is a usual classical
presentation of the motionless charge surrounded by the electric
field. The mass $m$ that enters {\it quantum} {\kern 0.5pt}
equation (4) is also a pure {\it classical} {\kern 0.5pt}
parameter. Hence even the most comprehensive quantum mechanical
description of the quantum system studied is only a
quasi-classical pattern.

Thus if one still wishes to remain devoted to orthodox quantum
mechanics, the fundamentals will be kept in the shade of its
statistical conformities.

On the other hand, since the tessellattice, or the real space is a
lattice of densely packed balls [45-47], all the phenomena that
can take place in the tessellated space a priori are
deterministic. In fact we do not need any necessity to introduce
long-range action because all kinds of interactions will now be
occur through superparticles, building blocks of the space. This
means that any actual microscopic theory should be specified by
short-range action, i.e. it should be characterized by carriers,
elementary excitations, or quasi-particles of the tessellated
space. In our theory such excitations were called 'inertons' [43].
It seems this appellation is best matched to their physical nature
because 'inertia' means a resistance to the motion. In fact any
particle moving in the tessellattice should undergoes collisions
on the side of coming cells, superparticles. Such an interaction
results in the creation of the tessellattice's excitations,
inertons, which therefore for ever become attached to particles.
Moreover, in paper [56] it was directly shown that a particle's
inertons carry the proper inert mass of the particle, $m_{{\kern
0.4pt} 0}/\sqrt{1-v_0^2/c^2}$. The value of mass of an inerton can
vary approximately from $10^{-85}$ to $10^{-45}$ kg [57].

\subsection{Wave-particle}

\hspace*{\parindent}  Classical mechanics is constructed in the
real space (i.e. in a 3D space or 4D space-time) where particles
-- material points -- are endowed with such measurable
characteristics as the position, velocity, momentum, and kinetic
energy. A classical wave is specified by measurable properties as
well, namely, the wavelength and frequency. In higher energy
physics particles do not demonstrate wave properties as well; they
seem material points. Thus, wave properties of particles, which
appear at low and intermediate energies, can indeed be associated
with an intervention of a subquantum medium, i.e. space, that
imposes the wave behavior on moving particles.

In 1924 de Broglie (see, e.g. Ref. [61]) formulating the Jakobi
theorem in the real space derived the de Maupertuis principle: A
real trajectory of a particle that moves from point $A$ to point
$B$ of the space is characterized by the minimum action
\begin{equation}
S=Et-m v{\kern 1pt} (\alpha x +\beta y + \sqrt{1-\alpha^2
-\beta^2} {\kern 1pt}z)
 \label{5}
\end{equation}
where is $E$  is the kinetic energy of a particle, $m$ its mass,
$v$ its velocity, $t$ is time, and $\alpha$ and $\beta$ are
direction cosines.

Then de Broglie derived the solution to the wave equation that
described a monochromatic wave spreading in an isotropic medium.
This allowed him to write the phase of the wave in the form
\begin{equation}
\varphi= \nu {\kern 1pt}t -\frac {1}{\lambda}{\kern 1pt} (\alpha x
+\beta y + \sqrt{1-\alpha^2 -\beta^2} {\kern 1pt}z)
 \label{6}
\end{equation}
where $\nu$ is the frequency of the wave and $\lambda$ its
wavelength. $\varphi$ is the total integral of the equation of
geometric optics, which takes account of the Fermat principle: A
real ray spreading from point $A$ to point $B$ is characterized by
the minimum phase.

Setting $\nu = E/h$ and then comparing expressions (5) and (6) de
Broglie came to his famous relationships for a canonical particle
\begin{equation}    E=h\nu \quad \ \ \
{\rm and} \ \ \ \quad \   \lambda =h/mv,
 \label{7}
\end{equation}
In expressions (7) parameters $E$ and $mv$ (the energy and the
momentum, respectively) belonged to the particle, but the
frequency $\nu$ and the wavelength $\lambda$ were characteristics
of a wave that should accompany and guide the particle at its
motion in the real space.

Note that the relationships (7) enable one readily to derive the
Schr\"odinger equation [61]. So, de Broglie's transparent idea
that a moving particle is accompanied by an actual wave did not
receive any further development.

The Schr\"odinger and Dirac formalisms say nothing about true
trajectories of the quantum system studied that is a direct
consequence of the probabilistic approach to the description of
quantum phenomena. Of course, one could use Feynman diagrams for
any entity, with their point-like particles and photons, all
having some absolute position and momentum. However, we cannot get
the true path. Instead we must draw infinitely many Feynman
diagrams and then calculate Feynman's path integrals, which make
it possible to find out only the most verisimilar trajectory of
the quantum system.

Once again, the inner reason is that conventional quantum
mechanics is developed in the phase space, but not in the real
one. Indeed, can one clarify the duality of a "wave-particle" in
the real space where only a particle and a wave can separately be
determined? The same is noted by Ligare and Olivery [62]: "It is
not always clear which aspects of classical wave behavior are
related in a fully quantum-mechanical treatment, or where to draw
the line between wave-like aspects and particle-like aspects and
how to justify the division".

Since in the author's approach  the real space is the
tessellattice formed by densely packed superparticles, a mechanics
of a particle should be different from that typical for classical
mechanics. The mechanics constructed in the tessellattice must
take into account the interaction of a moving particle with the
surrounding lattice. A detailed theory of the motion of a particle
interacted with superparticles of the tessellated space was
constructed in Refs. [43,44,50]. It has been argued that a
deformation coat, or a crystallite, is formed around a created
particle, which is identical with the deformation coat that is
formed around a foreign particle in the crystal lattice. The size
of the crystallite is associated with the Compton wavelength of
the particle, $\lambda_{{\kern 1pt}\rm Com} =h/mc$ and the role of
the crystallite is to shield the particle from the degenerate
space. The mechanics constructed is exemplified by elementary
excitations of the surrounding space, i.e. inertons, which
accompany the moving particle. The Lagrangian of a moving particle
has the form (simplified here)
\begin{equation}
L= \tfrac 12 {\kern 2pt} g_{ij} {\dot X}^i{\dot X}^j +  \tfrac 12
\sum\limits_{l=1}^N {\tilde g}_{ij}^{(l)} {\dot x}^i_{(l)}{\dot
x}^j_{(l)} - \sum\limits_{l=1}^N \frac{\pi}{T_l}{\kern 2pt}
\delta_{ij}{\kern 1pt} \big[ X^i \dot x_{(l)}^j + v {\kern 1pt}
x_{(l)}^j \big].
 \label{8}
\end{equation}
Here the first term describes the kinetic energy of the particle,
the second term depicts the motion of the ensemble of $N$ inertons
and the third term characterizes the interaction between the
particle and the ensemble.

In the so-called relativistic case, when the particle's velocity
$v$ approaches to the velocity of light $c$, the Lagrangian is
chosen in the classical form
\begin{equation}
L=-m_0c^2 \sqrt{1-v^2/c^2}
 \label{9}
\end{equation}
in which, however, the following transformation is made
\begin{equation}
v \longrightarrow \big[ g_{ij} {\dot X}^i{\dot X}^j + f(X,x, v,
\dot x) \big] /g
 \label{10}
\end{equation}
where the function $f$ includes terms analogous to the second and
third terms of expression (8).

Equations of motion of the particle and its inertons have been
studied in detail. It has been shown that the motion in fact is
marked by de Broglie relationships (7). Moreover, the inner
meaning of all the parameters has been clarified. In particular,
the particle's de Broglie wavelength represents the amplitude of
spatial oscillations of the particle. The particle emits inertons
running odd sections $\lambda/2$ of its path and its velocity
gradually decreases from $v$ to 0. During even sections
$\lambda/2$ the particle re-absorbs inertons again and its
velocity is restored to the value of $v$. The frequency $\nu =
1/2T$ where $T$ is the period of collisions of the particle with
the center-of-mass of its inerton cloud.

The particle's inerton cloud oscillates around the particle with
the same frequency $\nu$ and the cloud's amplitude $\Lambda$
satisfies expression
\begin{equation}
\Lambda  \simeq \lambda {\kern 2pt} c/v
 \label{11}
\end{equation}
where $c$ is the velocity of inertons, which is of the order or
over the velocity of light. The theory constructed in some aspects
is similar to the kinetic theory of a system of two linked objects
that are characterized by free path lengths $\lambda$ and
$\Lambda$.

The availability of the oscillating inerton cloud that accompanies
a moving particle automatically means that the particle indeed
goes with a peculiar real wave. Since the particle is always found
in the center of the system 'particle--inerton cloud', the system
as a whole may be treated as a wave travelling along a given
direction. The range of the space covered by the particle's
inerton cloud determines an area of employing of the wave
$\psi$-function formalism. So the double solution theory approach
allows us to fill the pure probabilistic interpretation of the
Schr\"odinger wave $\psi$-function with the concrete physical
contents described above. The problem of the measurement of an
abstract wave $\psi$-function when it collapses to a measurable
actual point now can easy be account for the re-absorption of
inertons by their particle.

\subsection{Matter waves}
\hspace*{\parindent}  Although in quantum mechanics the behavior
of particles are described by pure formal statistical formalism,
canonical particles, nevertheless, demonstrate properties of real
waves, i.e. the matter waves, which received empirical
confirmation in the diffraction experiments. Therefore, particles
in fact possess real wave properties predicted by de Broglie,
which automatically implies  that the pure probabilistic
interpretation of the $\psi$-function is not complete.

Briner et al. [63] published an experimental work entitled
"Looking at Electronic Wave Functions on Metal Surfaces", in which
they demonstrated the colored spherical and elliptical figures,
which the authors called "the images of $\psi$ wave functions of
electrons". Virtually they gave the evidence that the electron is
not a point-like object, though the high energy physics asserts
that it is a point object with the size no larger than $10^{-18}$
m. Thus they fixed an actual perturbation of the space around an
electron in the metal! The authors subconsciously rose against the
probabilistic interpretation of the $\psi$ wave accepted by the
Copenhagen School concept and, moreover, they practically proved
the fallaciousness of the statement of the concept. Consequently,
the experimental data point to the fact that the wave
$\psi$-function is not abstract but measurable matter.

Evidently, inertons as a substructure of the matter waves have
already received an implicit support among researchers [63]. We
shall discuss other manifestations of inertons in Section 3.

\subsection{Lorentz invariance}

\hspace*{\parindent} All correct theories should be Lorentz
invariant, i.e. they and Einstein's special relativity should
agree (see, e.g. Ref. [64]). Nevertheless, the Schr\"odinger
equation is not Lorentz invariant but it perfectly describes
quantum phenomena and we trust wholly the results derived from the
equation. How is it possible?

It seems that the disagreement between the strong theoretical
conclusion and the experimental veracity is hidden in the
statistical approach to the Schr\"odinger formalism. In papers
[43,44] the Schr\"odinger equation was derived from deeper first
principles that in fact removed a very unpleasant conflict that so
far took place between nonrelativistic quantum mechanics and
special relativity: Unlike the traditional presentation, the
Schr\"odinger equation gained in paper [44] is Lorentz invariant
owing to the invariant time entered the equation. Besides, it has
been shown in Refs. [44,56] that the tessellated space contracts a
moving object and its cloud of inertons in accord to the formalism
of special relativity (see also Ref. [46]), i.e. by the factor of
$\sqrt{1-v^2/c^2}$.

\subsection{Unification of Schr\"odinger and Dirac formalisms}

\hspace*{\parindent} Why Schr\"odinger's and Dirac's approaches
are so dissimilar? Why is an intermediate approach lacking?  It
seems that this issue has never been raised by researchers so far
at all. Nevertheless the problem is very serious and it must be
resolved.

In orthodox quantum mechanics there is no singled valued
parameters $E$ and $\nu$ in the expression $E = h \nu$ applied to
a moving canonical particle. In one case $E=\frac 12 {\kern 1pt}
m_0{\kern 1pt}v^2$ (see, e.g. Schiff [60], p. 33 ), and in the
other one $E = m_0 c^{\kern 1pt 2}/\sqrt{1-v^2/c^2}$ (see, e.g.
Schiff [60], p. 364). Which is true?

The problem has been studied by the author in paper [50]. To
answer the question let us consider the three relationships below,
namely: de Broglie's  $\lambda = h / m v$, Compton's
$\lambda_{{\kern 1pt}\rm Com} = h /m c$, and  $\Lambda \simeq
\lambda {\kern 2pt} c/v$ (11). This allows relation (11) to
rewrite as follows
\begin{equation}
\Lambda \simeq \lambda_{{\kern 1pt}\rm Com} {\kern 2pt} \frac
{c^2}{v^2}.
 \label{12}
\end{equation}

It has been shown in Refs. [44,50] that a moving particle
periodically passes its kinetic energy $\tfrac 12 {\kern 1pt}
m_0v^2/\sqrt{1-v^2/c^2}$ on to the particle's inerton cloud. That
is why, as follows from relation (12) when the velocity $v$ of a
particle satisfies the inequality $v << c$, the inerton cloud that
guides the particle carries the particle's kinetic energy $E=\frac
12 {\kern 1pt} m_0{\kern 1pt}v^2$. It is the energy that is
measured by the tool. At a distance about $\Lambda$ from the
particle its inerton cloud undergoes obstacles and passes the
corresponding information to the particle. This is the typical de
Broglie's "motion by guidance" and the utilization of the
Schr\"odinger formalism is quite correct in this situation.

When $v \rightarrow c$, the inerton cloud becomes virtually closed
inside of the particle's crystallite whose size is determined by
the Compton wavelength $\lambda_{\rm {\kern 1pt} Com}$. The total
energy of the crystallite coincides with the total energy of the
particle, $E = m_0 c^{\kern 1pt 2}/\sqrt{1-v^2/c^2}$, and if so
the tool will measure this energy. In such a manner when $v
\rightarrow c$ the energy of the particle at rest $m_0c^2$
explicitly reveals itself at the measurement and, therefore, in
this approximation the Schr\"odinger formalism fails and the Dirac
formalism becomes effective.

\subsection{Spin}

\hspace*{\parindent} What is spin? It is one more mystery of the
microworld. In quantum mechanics spin is perceived to be a certain
inner property of canonical particles. Quantum field theories
define spin as an "inseparable and invariable property of a
particle" (see e.g. Ref. [64], p. 17). That is all.

As a rule the notion of spin of a particle is associated with an
intrinsic particle motion. Several tens of works have been devoted
to the spin problem. Major of them is reviewed in recent author's
papers [50,52]. Main ideas of the works quoted there were reduced
to a moving particle that was surrounded by a wave, or a small
massless particle, or an ensemble of small massless particles,
which engaged in a circular motion.

Of course, it seems quite reasonable to assume that spin reflects
some kind of proper rotation of the particle. However, canonical
particles possess also electrodynamic properties and the operation
{\it rotor} is the principal characteristic of the particle's
electromagnetic field. Since quantum electrodynamics and quantum
mechanics of a particle must be in accord, the idea of rotation
regarding the notion of the particle's spin should be abandoned.

In the author's concept particles are determined as spatial
objects in the real space, which in fact makes it possible to
investigate the notion of spin in detail. In this case along with
an oscillating rectilinear motion, the particle may undergo also
some kind of an inner pulsation, like a drop. Two possible
pulsations of a particle either along its velocity vector or
diametrically opposite to it have been associated with the two
possible projections of the particle spin, i.e., the two own
pulsations of the particle in the real space are exhibited by two
so-called spin-1/2 projections, $\pm {\kern 1pt}\hbar/2$, in the
phase space [50]. Any spin bigger 1/2 is the property of a
composite quantum system.

The Pauli principle makes allowance for the spin "polarization" of
inerton clouds of two interacting particles. If inertons of the
two particle transfer the same projection of spin determined
above, the interaction will be repulsive; if spins of the
particles are opposite, the interaction will be attractive.

\subsection{Heisenberg's uncertainties}

\hspace*{\parindent}  Although there are Heisenberg's
uncertainties for the coordinate and momentum and the energy and
time of a particle,
\begin{equation}
\Delta x {\kern 1pt}\Delta p \geq \hbar, \ \ \ \ \ \Delta E {\kern
1pt} \Delta t \geq \hbar,
 \label{13}
\end{equation}
we are not able to write any similar relation for the particle
mass, which should also be fuzzy in a undetermined volume, the
same as the particle itself (the mass must follow the particle!),
as the probabilistic formalism prescribes.

De Broglie [65,66] studied the problem of the mass behavior and
came to the conclusion that the dynamics of particles had the
characteristics of the dynamics of the particles with a variable
proper mass. He was the first to indicate that the corpuscle
dynamics was the basis for the wave mechanics.  With the
variational principle, he obtained and studied the equations of
motion of a massive point reasoning from the typical Lagrangian
\begin{equation}
L = -M_0c^{\kern 1pt 2}\surd\overline{1-v^2/c^{\kern 1pt 2}}
\label{14}
\end{equation}
in which the velocity $v$ of the point and the velocity of light
$c$ were constant along a path. De Broglie's pioneer research
allows one to suggest that a real wave, which indeed has to
accompany the moving particle, must complement the deficient value
of the momentum and the energy of the particle. Then, say, we know
the momentum and the energy, but have uncertainties in coordinate
and time. If we assume the existence of an actual wave that
travels in the space together with the particle, we can readily
propose that the particle is entrained by the given wave and,
therefore, position and time of the particle become in fact
undetermined in a concrete point as they become functions of the
travelling wave.

The uncertainty principle is a direct consequence of the
probabilistic approach to quantum phenomena when only one of two
subsystems is taken into account, namely, we treat the behavior
only a particle, but totally ignore its inerton cloud that
accompanies the particle.

It seems that in some situations the uncertainties are sound not
universally true. For instance, Gong [67] has recently re-analyzed
two well-known ideal experiments: (i) Heisenberg's $\gamma$-ray
microscope experiment and (ii) the single slit diffraction
experiment. He has shown that in the case (i) the uncertainty
principle cannot be employed for a quantum system if its size is
equal or large than the resolving limit $\Delta x_{\rm mic}$ of
the microscope. In the situation (ii) if a particle has a certain
position in the slit, the uncertain quantity of the position is
also wanting. Hence in both cases the relation $\Delta x {\kern
1pt}\Delta p \sim 0$ is held, which is incompatible with the
prediction (13).

A conclusion can be drawn that the uncertainty principle is not
universal. It should be applied with a great caution. The reason
is that the nonlinear behavior of a quantum system would stem from
a subtle interaction with the environment, which is discussed
below.

\section{Field generated by motion}

\subsection{Experimental evidence}

\hspace*{\parindent} Classical mechanics itself has evidence that
the space should be treated as a substrate, though such direct
demonstrations as inert forces and the centrifugal force are not
yet taken into account in quantum and gravitational physics.

At the same time quantum theories that describe electrodynamics,
weak and strong interactions are characterized by their own
carriers, quasi-particles or particles: photons, $W^{\pm}$- and
$Z^0$-bosons, and gluons, respectively. In this paper we have
argued that quantum mechanics being constructed in the real space
immediately gives rise to its proper carriers, namely, inertons.
We have already talked about experimental evidence presented by
Baurov [40] and Briner et al. [63]. Here we would like to inform
about other recent data. Benford [68] could record radiation of a
unknown nature from a ferrite disk revolving on its axis.
Urutskoev and co-workers [69] registered a new "strange" radiation
as well when they studied transmutation of chemical elements in
foils during electric discharge -- the effect that is still
considered as completely impossible among the majority of nuclear
physicists.

The experimental verification of submicroscopic quantum mechanics,
namely, that moving objects are accompanied by inertons,
elementary excitations of the space, has been carried out in
papers [54,70,71]. In work [54] we started from a hypothesis that
in condensed media inerton clouds of separate entities should
overlap forming the entire inerton field that should be quantized
the same as the phonon field. In a solid, the force matrix $W$
determines branches of the solid's acoustic vibrations. We
supposed that the force matrix should include in addition to the
phonon term also the term associated with the overlapping of
inerton clouds; so, the force matrix transformed to $W = W_{\rm
acust}+ W_{\rm inert}$. Therefore an outside inerton field would
be able to influence the solid in the same way as the acoustic
field does. Since the Earth is a power source of the inerton
radiation. We in fact could fix changes in the fine morphological
structure of samples studied, Figure 1 (experimental details see
in Ref. [54]). The device that measures the inerton radiation has
recently been constructed by my colleagues, Figure 2.

In paper [70] the anomalous photoelectric effect occurring under
strong irradiation was examined from the submicroscopic
standpoint. The phenomenon, in essence, is this: these are
electrons' inerton clouds which absorb photons of an incident
laser beam, because the cross-section of an electron's inerton
cloud $\sigma_{\rm cloud}$ much exceeds that of an atom
$\sigma_{\rm atom} \sim 10^{-2}$ nm$^2$ (the actual electron's
cross-section satisfies inequalities $\lambda^2 < \sigma_{\rm
cloud} < \Lambda^2$). It has been shown that the multiphoton
theory that is still employed by the majority of specialists in
quantum optics since the mid-1960s is wrong as the photoelectric
effect studied has a linear dependence on the intensity of light,
though the multiphoton theory is strongly nonlinear theory; this
also has long been stressed by Panarella [71]. The result [70] is
supported by comparison with a great numbers of experimental data.

In paper [72] it is shown that in the KIO$_3 \cdot$HIO$_3$ crystal
hydrogen atoms co-operate in peculiar clusters in which, however,
the hydrogen atoms do not move from their equilibrium positions,
but become to vibrate synchronously. The interaction between the
hydrogen atoms is associated with the overlapping of their
inertons. The exchange by inertons results in the oscillation of
mass of hydrogen atoms, which manifested itself in the IR spectra
analysed.
\begin{figure}
\begin{center}
\includegraphics[scale=0.8]{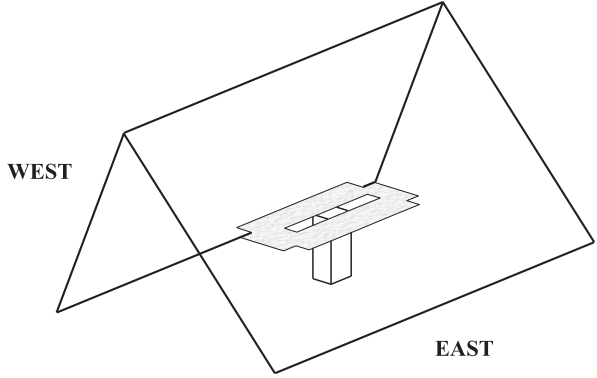}
{\caption{\small{Resonator of the Earth inerton field made of two
transparent organic glass plates and the sample (a razor blade)
whose fine morphologic structure changed under the Earth inerton
radiation, which was fixed by the electron microscope [54] (for
further reading about the phenomenon see Ref.
[73].)}}\label{Figure1}}
\end{center}
\end{figure}
\begin{figure}
\begin{center}
\includegraphics[scale=0.5]{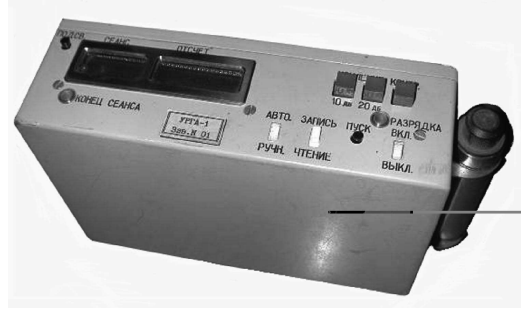}
\caption{\small{Device "URGA-1" (the Universal Registrar of
Geological Anomalies) that measures the inerton radiation, in
particular, in pyramid-like constructions with an effective size
over 1 m.}} \label{Figure 2}
\end{center}
\end{figure}

\subsection{Prospects for quantum physics }

\hspace*{\parindent} Since photons  and inertons are two major
quasi-particles, which are excited in the tessellated space [74],
we would build up a submicroscopic theory of the diffraction
phenomenon both for particles and photons, which today is still
limited by the geometrical optics approximation. The theory will
also elucidate a microscopic origin of the phenomenon of
dispersion of light and, moreover, it will show that just inertons
-- carriers of the matter waves -- play the fundamental role in
the diffraction/dispersion phenomenon.

The said is not a fantasy by the author. A submicroscopic theory
of the phenomenon of the diffraction of photons/particles promises
new interesting results that would be very considerable for
applied physics. Indeed, a mechanism of the diffractionless of
single photons revealed by Panarella (see, e.g. his review article
[75]) about 20 years ago remains completely unclear. The behavior
of photons was typical for that of classical corpuscles: single
photons passed through a pinhole and did not form any fringe on
the target. This would mean that the diffraction pattern is formed
only due to the interaction of each travelling photon with the
surrounding clouds of inertons, which ceaseless oscillate around
their entities. The diffractionless would mean a peculiar photon
channelling when a travelling photon is not scattered by
surrounding inerton clouds of the matter. At what conditions is it
possible? The answer will shed light on other quantum phenomena
such as entanglement, teleportation, etc. Thus the inerton field
is capable to play a role of the control field that will replace
an indeterministic 'randomness of the quantum world'.

Nakhmanson [16] notes that the idea of informational experiments
with particles has never been publicly discussed. Then he
continues that if particles have consciousness, they may receive a
signal and then will transfer it to matter. Submicroscopic quantum
mechanics constructed in the real space completely supports this
idea, because the particle's inerton cloud plays the role of a
peculiar 'consciousness' of the particle, which was hypothesized
in Ref. [16].

As the control field, a flow of inertons would provide for
controlled low energy nuclear reactions like those that were
revealed by Benford [68] and Urutskoev et al. [69], though in
their experiments they only claimed about a new "strange"
radiation. In the recent research conducted by our team we have
obtained much more remarkable results in this area as long as we
followed the submicroscopic theory stated above.

\section{Concluding remarks}

\hspace*{\parindent} Thus, if we turn from the statistical
standpoint on the nature of everything, we arrive at the very new
pattern of the physical world described above. New peculiarities
and links come to light. From the viewpoint of submicroscopic
determinism many urgent problems of contemporary physics abruptly
lose their supreme meaning, because these problems appear now as
typical phenomenological issues whose decision is found at a
deeper level of our knowledge.

For instance, Arunasalam [59] has recently discussed the views on
the fundamental problems expressed by different sets of famous
physicists (Einstein, Dirac, Feynman, Pauli, Bethe and others) and
shown that the views are in sharp contrast: covariance versus
invariance, relativistic versus nonrelativistic electron theories,
etc. The problems indeed are considerably important. However, if
we are resting on  deeper, i.e. deterministic first principles,
such problems as charge conjugation $\mathbf{C}$ and
$\mathbf{CPT}$ violation seem not determined owing to vagueness of
the notions {\it charge}, {\it space}, and {\it time} in the
framework of the theory that tries to resolve the problems.

From the viewpoint of the theory of space constructed in works
[45-47] and submicroscopic quantum mechanics discussed above
modern theories on particle physics do not show up as fundamental.
In the future, without doubt, they will be re-analysed on the
basis of the submicroscopic concept. Similar problems remain to
face gravitational physics. For example, the problem of
gravitational waves seems very farfetched. Gravitational waves are
not a realistic solution to the Einstein equations, as has been
shown by Loinger [76], and they are forbidden by submicroscopic
quantum mechanics developed in the real space [54].

Consequently, further sophisticated study of the constitution of
the real space and the deriving matter and physics laws from the
space are the shortest road to the progress of science and the
advanced technology.

\end{document}